\documentclass[10pt,journal,compsoc]{IEEEtran}
\usepackage{epsfig,endnotes,url,color,subfig}
\usepackage{graphicx} 
\usepackage{xspace} 
\usepackage{wrapfig} 
\usepackage{amsmath,amssymb} 

\newcommand{\micro}{\ensuremath{\mu}}

\usepackage [linesnumbered,lined,ruled,commentsnumbered]{algorithm2e}
\usepackage {algpseudocode}
\usepackage{color, colortbl}

\newcommand{\projecttitle}{\textsc{EVECTOR}\xspace}

\newcommand{\eat}[1]{}

\newcommand{\myparagraph}[1]{\smallskip \noindent{\bf {#1}.}}

\newcommand{\out}[1] {}

\newcounter{codeLineCntr}

\reversemarginpar
\newif\ifnotes
\notestrue

\newcommand{\punt}[1]{}

\renewcommand{\eqref}[1]{Equation~(\ref{#1})}

\newcommand{\proc}[1]{\ifmmode\mbox{\textsc{#1}}\else\textsc{#1}\fi}

  \newcommand{\func}[1]{\ifmmode\mathrm{#1}\else\textrm{#1}fi} %

\newcounter{remark}[section]

\begin{document}

\title{\projecttitle: An orchestrator for analysing attacks in electric vehicles charging system} 

\author{Devki Nandan Jha,
        Tomasz Szydlo,
        Nima Valizadeh,
        Ringo Sham,
        Aleksandra Edwards,
        Amrit Kumar,
        Amanjot Kaur,
        Bo Wei,
        Vijay Kumar,
        Kai Li Lim,
        Rajiv~Ranjan,
        Omer Rana

\IEEEcompsocitemizethanks{
	\IEEEcompsocthanksitem D. N. Jha, T. Szydlo, R. Sham, A. Kumar, B. Wei and R. Ranjan are with Newcastle University, UK. E-mail: \{Dev.Jha, Tomasz.Szydlo, Ringo.Sham, Amrit.Kumar, Bo.Wei, Raj.Ranjan\}@ncl.ac.uk
	\IEEEcompsocthanksitem N. Valizadeh, A. Edwards, A. Kaur, V. Kumar, O. F. Rana is with Cardiff University, UK, E-mail: \{ValizadehN, EdwardsAI, KaurA7, Kumarv14, RanaOF \}@cardiff.ac.uk
     \IEEEcompsocthanksitem  K.L. Lim is with University of Queensland, Australia. Email: Kaili.lim@uq.edu.au
}
}

\IEEEtitleabstractindextext{

\begin{abstract}
Electric Vehicle (EV) charging infrastructure is critical for the widespread adoption of EVs, ensuring efficient and secure charging processes. Evaluating the security and performance of EV charging systems in real-world infrastructure poses significant challenges due to the diversity of information exchange between vehicles and charging stations/Electric Vehicle Supply Equipment (EVSE), including complex network protocols, scale of deployment and a variety of potential threats. Existing simulation frameworks are unable to handle complex security scenarios across these differing data exchange protocols. In this paper, we propose a novel EV orchestration framework: \projecttitle, which addresses the limitations of existing simulation systems by enabling both quantitative and qualitative analyses of EV charging scenarios. \projecttitle also provides a flexible attack orchestrator to simulate realistic attack behaviours on EV charging infrastructure. We validate the \projecttitle framework through two case studies: (a) cyber-physical attacks such as {\it broken wire}; and (b) cyber-specific attacks such as {\it frame fuzzification}. The case studies highlight the effectiveness of \projecttitle in providing deeper insights into the security and performance of EV charging systems.
\end{abstract}

\begin{IEEEkeywords}
Electric Vehicles, EV Charging Protocols, EV Simulators, Security Threats, Orchestration
\end{IEEEkeywords}}

\maketitle
 
\section{I\lowercase{ntroduction}}
\label{sec:introduction}

Electric Vehicles (EVs) are now available from (virtually) every vehicle manufacturer -- offering a sustainable and environment-friendly alternative to traditional diesel and petrol vehicles. More than $3$ million EVs were sold in the first quarter of $2024$, an increase of $21$\% compared to $2023$~\cite{ev_trends}. As the number of EVs on the road increases, the infrastructure supporting them becomes crucial. The widespread adoption of EVs relies on the availability and reliability of a robust charging infrastructure -- ranging from home chargers to complex public and fast-charging networks designed to accommodate various vehicle types and charging standards~\cite{narasipuram2021technological,mastoi2022depth}. EV charging involves physical cable connections, integrated with communication protocols and software management systems~\cite{acharige2023review} -- representing a complex, distributed cyber-physical system. Ensuring the seamless integration of components is essential for the reliable functioning of the EV ecosystem.

The increasing reliance on the charging infrastructure brings numerous risks and threats to the entire EV ecosystem~\cite{antoun2020detailed,johnson2022cybersecurity}. An attacker can take advantage of the vulnerabilities in charging protocols and software management systems to launch various types of attacks, including unauthorised access to Electric Vehicle Supply Equipment (EVSE), disruption of charging services, data breaches and possibility of damaging a vehicle’s battery or other critical physical systems. Additionally, these threats can undermine confidence in EV technology, thus hindering their widespread adoption~\cite{ev_adopt}.

The early detection of threats to EV charging infrastructure is critical for ensuring the safety and reliability of the EV ecosystem. Identifying suspicious activities, such as unauthorised access requests, abnormal communication patterns, or system failures (malfunction), allows administrators to act proactively to prevent potential attacks. However, detecting these threats early has significant challenges, which include the complexity and diversity of the EV charging system, vulnerability in physical and software components, integrating emerging systems with legacy systems, and balancing security methods with usability~\cite{aljohani2024comprehensive,ye2020cyber}. Existing EV charging security research includes encryption and authentication mechanisms~\cite{hettiarachchi2024survey,nguyen2023authentication,parameswarath2022user}, use of distributed ledger technologies~\cite{rana2024enhancing,zhang2021privacy}, and AI-based anomaly detection techniques~\cite{hussain2024anomaly,jahangir2024charge}. 

To understand the likely impact of attacks on EV charging, we propose an orchestration environment able to combine multiple simulation systems. Each simulation system captures particular types of attacks, and there is no single simulator able to provide end-to-end simulation of components that make up an EV ecosystem~\cite{kellner1999software}. Various simulators/ frameworks are available for EV charging and management, each handling specific aspects of the ecosystem, e.g. simulators such as EV2gym~\cite{orfanoudakis2024ev2gym} and ACNSim~\cite{lee2019acnsim} which focus on EV charging, scheduling and energy distribution. On the other hand, ISO implementation~\cite{iso} and EVerest~\cite{everest} support EV charging protocols like ISO~15118 and Open Charge Point Protocol (OCPP). Security-oriented frameworks such as EVShield~\cite{evshield} help analyse cyber threats and vulnerabilities in charging infrastructure. Existing simulators however lack the capability to comprehensively orchestrate EV charging scenarios, protocols, and attack vectors in an integrated manner, limiting their effectiveness in holistic security and operational analysis. A key observation underpinning \projecttitle is that no one simulator can capture the diversity of threats within an EV ecosystem. There is a need for an orchestration mechanism to integrate multiple context-specific simulators and the outcome of these integrated for subsequent analysis.  

\subsection{Challenges}
Implementing a robust simulation orchestrator for EV charging is challenging. To be specific, the main challenges are as follows:

\myparagraph{Complex EV ecosystem} The EV charging ecosystem can include various types of EVSE (Level 1, Level 2, and DC fast chargers), different EV models and multiple communication protocols (e.g. ISO 15118), network protocol (e.g. OCPP, OCPI), physical connectors and charging standards (e.g. IEC 61851 (Type 2), SAE J1772 (Type 1), SAE J3400 (NACS), CCS, etc.) in different functional categories. Moreover, an EV can have differing battery capacity, charging rates and compatibility with charging standards. Similarly, an EVSE may support unique communication interfaces, power modes and software systems. Identifying the granularity at which this variation can be represented in a simulated environment requires an understanding of each component and its interactions -- and potential expected outcome from the simulation. 

\myparagraph{Variety of Threats} The threat landscape of the EV ecosystem is highly dynamic, with new vulnerabilities and attacks emerging regularly.  Additionally, the protocols and software stack within the EV ecosystem are continuously being updated, which can introduce new types of attacks and expose previously unknown security gaps. Each attack method can exploit different aspects of the EV ecosystem. It is necessary to understand how these attacks operate and the components that they impact.

\myparagraph{Attack Explainability} Attack behaviours can involve protocol exploitation to physical tampering, requiring a simulator to trace the root cause and impact. Conversely, the high volume of data generated by simulation can hide critical attack indicators, requiring effective visualisation and logging mechanisms. Balancing analysis with real-time detection makes it difficult to fully support explainability, as security insights must be both interpretable for analysts and actionable for automated defence.

\myparagraph{Real-time Simulation Needs} Some aspects of EV charging security, such as real-time attack detection and response, requires simulation to operate in real-time -- requiring significant computational resources.
    
\myparagraph{Data Availability} Creating a realistic simulation relies on access to high-quality data. This data can include EV charging behaviours and network traffic patterns. Acquiring such data is often challenging, as a number of EV charging systems are proprietary, with manufacturers and service providers resistant to sharing detailed data. Furthermore, data on cyber attacks is sensitive and may not be (easily) available. 

Given these challenges, a comprehensive and flexible orchestrator is needed to seamlessly integrate EV charging scenarios, communication protocols, and potential attack vectors. Such an orchestrator should support both realistic EV charging behaviours and associated security assessments -- enabling the execution of diverse user and attack scenarios. This paper presents \projecttitle, an \emph{E}lectric \emph{VE}hicle \emph{C}harging A\emph{T}tack \emph{OR}chestrator to address the challenges discussed above. \projecttitle is designed to provide a modular framework that supports multiple EV charging protocols, dynamic attack emulation, and scalable simulation. 

\subsection{Contribution}

\begin{itemize}
    \item The design and implementation of \projecttitle is described, which combines existing open-source EV charging simulators, including Everest and ISO 15118, to support simulation and analysis of a holistic EV charging ecosystem. 
    \item \projecttitle provides an attack orchestrator to support attack generation on the EV charging environment. \projecttitle also captures system logs  encoded using a pre-defined schema structure, which can facilitate further analysis.
    \item We evaluate \projecttitle on both cyber-physical and cyber-only attack scenarios: broken wire and data frame fuzzification attacks. The outcome can be used to assess vulnerabilities and impact on EV charging ecosystem.
\end{itemize}

\myparagraph{Outline} 
The remainder of the paper is as follows. The background along with recent, related work is presented in Section \ref{sec:related}. The system architecture of \projecttitle is given in Section \ref{sec:architecture}, specifying the components in the architecture and their interactions. The evaluation of \projecttitle on multiple scenarios is presented in Section~\ref{sec:evaluation}. We provide a discussion highlighting the limitations of our
work in Section \ref{sec:discussion}, with conclusions in Section \ref{sec:conclusion}. 
\section{Background and Related work}
\label{sec:related} 

\subsection{EV Charging Ecosystem}

\begin{figure}[t]
    \centering
    \includegraphics[width=\linewidth]{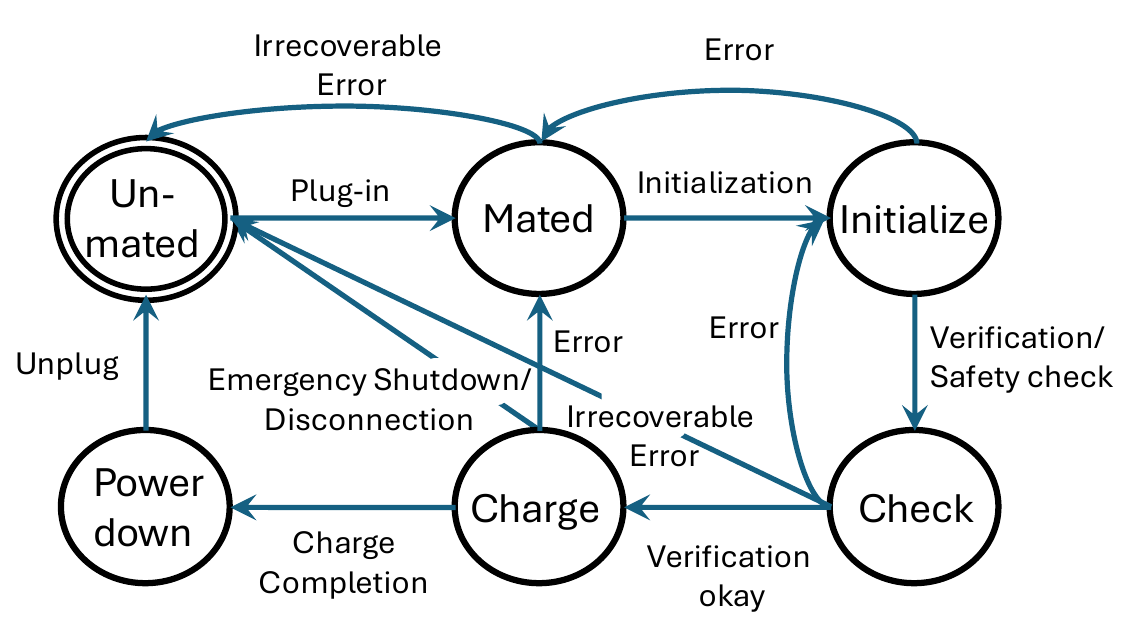}
    \caption{EV charging states}
    \label{fig:EV-states}
\end{figure}

The EV charging process involves a sequence of well-defined steps to ensure safe, efficient, and reliable energy transfer between the EV and the EVSE. 
Figure \ref{fig:EV-states} shows the various steps in the EV charging process and their connections. An explanation of each step, including the protocols involved, is given below.

\emph{1. Unmated State:} In the unmated state, an EV and EVSE are not physically connected. During this phase, the vehicle is parked near the EVSE, and the driver prepares for the charging session. No communication occurs at this stage. However, some charging systems with data exchange enabled may broadcast availability and identification information. In this case, EVSE may use OCPP to communicate status updates, such as heartbeat signals, to the backend management system.

\emph{2. Mated State:} The mated state begins when the charging cable is physically attached to the EV. A mechanical and electrical connection is established between the EV and EVSE. At this stage, the connection is verified to ensure safety. ISO 15118 comes into play to initiate communication between the EV and EVSE. The ISO 15118 protocol facilitates the handshake process to confirm compatibility, charging mode and initial configuration settings.

However, faults such as improper cable attachment or tampered connectors can occur here. A faulty vehicle, an EVSE or a physical tampering attack that damages the charging plug or port might prevent the system from transitioning to the initialisation state. To avoid this, the system may remain in the mated state, or in cases of persistent or irrecoverable issues, revert to the unmated state to prevent further damage.

\emph{3. Initialize State:} Once the physical connection is confirmed, the initialisation phase begins. The EV and the EVSE exchange critical information, such as the vehicle’s state of charge (SoC), battery capacity, and required energy levels. This ensures that the charging session is customised for the specific needs of the vehicle. For advanced vehicle-to-grid (V2G) capabilities, which require bidirectional energy transfer and dynamic grid communication, ISO 15118‑20 is employed to manage the necessary protocols and secure data exchange.
Additionally, OCPP facilitates communication between the EVSE and the backend system, allowing for user authentication, payment verification, and session authorisation.

An attack can also happen at this stage -- e.g. xommunication spoofing, man-in-the-middle (MITM) attacks or software bugs can disrupt this phase~\cite{ghafouri2022coordinated,hamdare2024mitm}. If an MITM attack occurs, falsifying authentication messages, the system may fail to authenticate and revert to the mated state.

\emph{4. Check State:} The check state ensures the readiness of all components involved in the charging process. Safety checks are conducted to confirm proper grounding, voltage levels and compatibility. The ISO 15118 protocol supports safety parameters, such as ensuring earthing and voltage checks are valid. In this stage, faults such as grounding issues, sensor malfunctions or false safety signal attacks e.g. injecting malicious data to mimic faults can cause the system to abort the session. In such cases, the system might transition back to the initialisation state to retry safety checks or terminate the session entirely by returning to the unmated state. In this case, ISO 15118 can provide fault diagnostics. Also, the OCPP protocol is used to communicate any detected issues to the backend system for logging and resolution.

\emph{5. Charge State:} The charge state involves the actual transfer of energy from the EVSE to the EV battery. During this phase, the power flow, voltage and temperature are continuously monitored, and adjustments are made based on the EV requirements and the EVSE capabilities. The real-time power delivery, checking the charging status and receipt of metering data
is maintained by the ISO 15118 protocol. The OCPP protocol keeps the backend system updated on the progress of charging sessions, energy consumption and time elapsed.

Faults such as sudden power surges, over-current conditions, or broken cable attacks (physical disconnection during charging) can force the system to halt charging~\cite{ghafouri2022coordinated,kohler2022brokenwire}. A broken cable attack or accidental physical disconnection during charging can force the system to halt energy transfer abruptly. The charging system would then detect the loss of connection and transition directly to the unmated state, skipping the standard power-down sequence. Similarly, a severe fault such as a fire hazard or power surge, might trigger an emergency shutdown protocol that cuts off power and disconnects the session, transitioning directly to the unmated state and protecting the equipment. The system may also move to the mated state in case a charging error such as voltage mismatch or overheating is detected. In this case, the system will stop energy transfer while maintaining the physical connection. 

\emph{6. Power Down:} As the charging session nears completion, the power-down phase ensures a safe and gradual reduction in energy transfer. The ISO 15118 protocol coordinates the controlled cessation of energy flow, ensuring all components are safely deactivated. OCPP communicates session termination to the backend system and handles payment processing or receipt generation. This step is critical to prevent abrupt power cuts, which can damage an EV battery or EVSE.

The final step returns the system to the unmated state. The EV is disconnected from the EVSE and the session ends. The EVSE becomes available for the next user.

\begin{figure}[t!]
    \centering
    \includegraphics[width=0.45\textwidth]{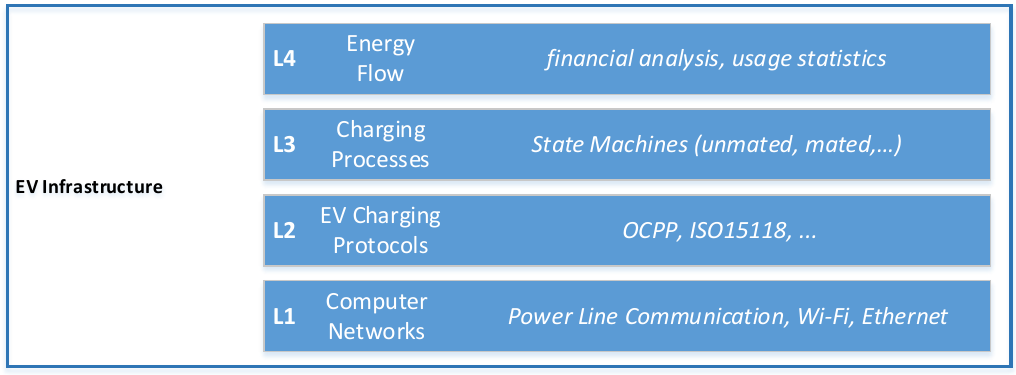}
    \caption{Layered architecture of EV charging infrastructure}
    \label{fig:layered-architecture}
\end{figure}

\subsection{Architecture for EV Charging}

The EV charging infrastructure can be divided into four layers depending on the network protocol being used, as shown in Figure \ref{fig:layered-architecture}.  Interactions between the different layers are given below.

\textit{A) Computer Network Layer (L1)} - This layer handles data exchange between EVs, EVSEs, backend servers and grid operators. It ensures secure connectivity using wired (Ethernet) and wireless (Wi-Fi, LTE, 5G) communication.


\textit{B) Protocol Layer (L2)} - This layer governs how EVs, EVSEs, and backend systems communicate, handling authentication, payment processing, and charging negotiations using standardised protocols such as ISO 15118 and OCPP. At this level, interoperability support is required to overcome communication problems between devices, especially those from different manufacturers.

\textit{C) Charging Layer (L3)} - This layer deals with the physical energy transfer process between the EV and the EVSE, i.e., start, duration and energy sent to the EV battery, ensuring power regulation, voltage/ current adjustments and safety mechanisms during charging are adhered to -- examples of protocols to support this include J1772 and IEC61851.

\textit{D) Energy Management Layer (L4)} - This layer connects EVSEs to the power grid, managing energy demand, balancing loads and ensuring grid stability. Protocols at this layer include IEC 61850 for EV charging for smart grid integration, V2G interactions and dynamic power allocation.

\subsection{Related Work}

There are multiple simulation frameworks available to support an EV charging environment. To gain a comprehensive understanding of EV charging, existing simulators and emulators can be broadly classified into three key.

\subsubsection{EV charging protocol simulators/emulators} This category of simulators/ emulators focuses on replicating communication standards to ensure seamless interactions between EVs and EVSE while maintaining compliance with industry protocols. These simulators assist in testing protocol adherence, validating secure charging transactions and ensuring interoperability.

Switch-EV~\cite{switchev} offers tools and platforms designed to implement and test V2G communication standards, with a strong focus on ISO 15118. One of its offerings, RISE V2G~\cite{risev2g} and Josev Community~\cite{Josev}, serves as a reference implementation for ISO 15118, providing features like Plug and Charge (PnC) for authentication and billing. OpenV2G~\cite{openv2g} is another open-source implementation of the ISO 15118 communication protocol. It supports custom testing scenarios and extensions for research environments. Another open-source EV charging software stack framework is EVerest~\cite{everest}, which currently supports ISO 15118 and OCPP. It is available as Docker containers and provided with a Node-Red user interface. Mobilityhouse OCPP Simulator~\cite{mobilityhouse} is a Python implementation of OCPP with support for versions 1.6 and 2.0 and uses the JSON version of the protocol. It also provides examples of how to implement an EVSE and client. Despite their advantages, these simulators/ emulators mainly focus on protocol behaviours and interoperability testing without incorporating dynamic attack simulations, security analysis, or integration with realistic EV charging scheduling and energy management scenarios.

\subsubsection{EV request-response simulators} This category includes simulators for EV transactions, charge scheduling, and power management, with the main focus on optimising energy distribution, forecasting charging demands and improving grid efficiency. 

V2G-Sim~\cite{v2gsim} offers rich EV modelling, including EV, EVSE, but is not open-source.  EVLibSim~\cite{rigas2018evlibsim} and EV-EcoSim~\cite{balogun2023ev} support various charging scenarios but are constrained by realistic EV behaviour data. OPEN~\cite{morstyn2020open} provides smart energy system simulations but features uniform EV models, making it unsuitable for evaluating attacks. ACN-Sim~\cite{lee2019acnsim} is an open-source simulator which utilises actual charging session data to model the behaviour of charging systems, including battery behaviour and charge scheduling. Chargym~\cite{karatzinis2022chargym} enables cost optimisation with EV simulation but oversimplifies EV behaviour and lacks any real-world data. EV2Gym~\cite{orfanoudakis2024ev2gym} is another simulator that models EV behavior based on real data from EVs, EVSEs and their interaction. EV2Gym is designed to assess the behaviour of smart charging algorithms within a standardised platform on a defined scale. While these simulators are effective in load balancing and power management, they lack integration with security mechanisms and do not account for protocol-level details or charging infrastructure vulnerabilities. Their primary focus remains on energy optimisation.

\subsubsection{EV attack handling simulators} Several studies have explored security vulnerabilities/ threats in EV charging infrastructure, particularly focusing on the ISO 15118 and OCPP protocols.  The Cyber-Energy Emulation (CEE) platform~\cite{sanghvi2021cybersecurity} simulates cyberattacks on EV chargers and their impact on the power grid, yet it lacks detailed insights into the specific attack methodologies and does not provide a structured approach for identifying vulnerabilities. Similarly, other work~\cite{mattepu2022ocss} analyses security flaws in OCPP 1.6 but does not explore the adaptability of the simulator to newer protocol versions or diverse system configurations. Sarieddine et al.~\cite{sarieddine2024uncovering} employ an OWASP-based testbed to identify critical vulnerabilities like man-in-the-middle attacks, and~\cite{sarieddine2023real} presents a real-time co-simulation testbed that integrates OCPP to analyse threats and power grid impact. EVShield~\cite{evshield} also helps analyse cyber threats and vulnerabilities in the charging infrastructure. Although these simulators provide valuable insights into security threats, they do not fully integrate with EV charging protocol validation, making it difficult to conduct end-to-end testing of EV charging environments.
 
\section{System architecture}
\label{sec:architecture}

\begin{figure}
    \centering
    \includegraphics[width=\linewidth]{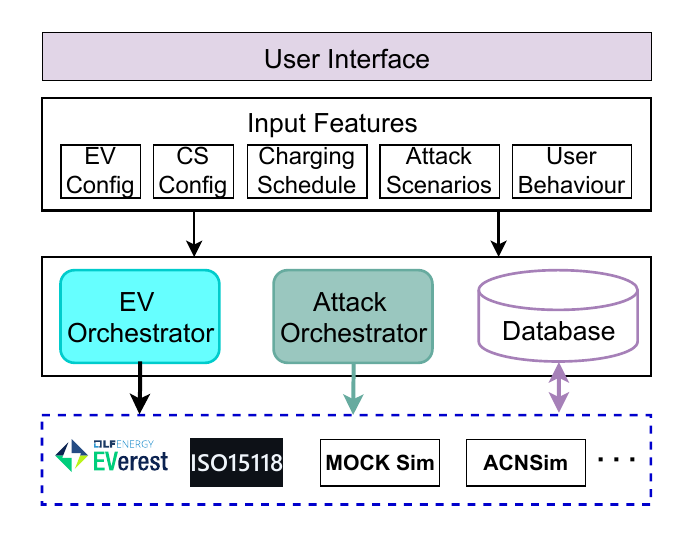}
    \caption{Proposed \projecttitle orchestrator framework}
    \label{fig:EV-arch}
\end{figure}

This section presents the architecture (illustrated in Figure \ref{fig:EV-arch}), design and implementation of \projecttitle.  

\subsection{\projecttitle architecture}\label{subsec:orchestrator}
\projecttitle is a modular and adaptable orchestration framework designed to replicate and analyse various aspects of the EV charging ecosystem. It consists of multiple layers, each responsible for a specific function, ensuring flexibility, scalability, and extensibility. This enables users to define EV charging scenarios, simulate different behaviours, introduce attack scenarios, and analyse the generated data for security insights. The following sections describe each layer in detail.

\myparagraph{A. User Interface}
This layer provides an interactive interface for users to configure and control the simulation environment, and an intuitive means for users to explore different EV charging scenarios. It allows users to input various parameters, including \emph{EV config} and \emph{EVSE (CS) config}. 
It also enables users to configure specific attack scenarios, such as \emph{Broken wire attack}, making it possible to study the impact of various cyber threats/ attacks on the EV charging infrastructure. 

\myparagraph{B. Input Layer} This layer serves as the primary configuration point, allowing users to define various aspects of the simulation. The main input features are given below.

1. \emph{EV Configuration}: Users can specify the number of EVs participating in the simulation, their battery capacities, charging requirements, SoC, and other operational parameters.

2. \emph{EVSE Configuration}: The system supports multiple EVSEs (charging stations), enabling users to define different station types, power ratings, charging protocols and availability.

3. \emph{Charging Schedule}: Users can input dynamic charging schedules that state when and how EVs should charge, incorporating real-world constraints such as peak and off-peak hours.

4. \emph{Attack Scenarios}: This feature enables users to simulate cybersecurity threats by selecting specific attack vectors (e.g., broken wire attack).

5. \emph{User Behaviour}: The orchestrator also allows the simulation of different user behaviours, such as preference for certain EVSEs, frequency of charging and interactions with the charging infrastructure.


\myparagraph{C. EV Orchestrator} This is the core component of \projecttitle, which acts as the engine for simulating the behaviour of EVs and the broader EV charging ecosystem. This component is responsible for replicating real-world charging interactions based on the configurations provided in the Input Layer. It integrates \emph{EV Behaviour Simulation}, \emph{EVSE Operations}, \emph{Charging Schedule} and \emph{Scenario} execution. It directly interacts with the \emph{Simulator Suite} to simulate various EVSE operations, including communication with EVs through protocols like ISO 15118 and OCPP. Based on predefined user scenarios, the orchestrator enables running specific test scenarios. Section \ref{sec:class-diagram} describes different classes and their interaction.
The modular nature of the EV Orchestrator allows it to be extended easily, supporting new EV and EVSE types, charging models, additional protocols, and emerging technologies in the EV ecosystem.

\myparagraph{D. Attack Orchestrator} The \emph{Attack Orchestrator} is another crucial component designed to inject security threats such as Broken wire attacks into the simulated EV environment. This component enables researchers and analysts to evaluate the resilience of EV charging infrastructures against cyber and physical attacks. The \emph{Attack Orchestrator} is a modular component. This paper details the design and implementation of 
\emph{Broken Cable attack} and \emph{Fuzzification attack}. By integrating real-time attack simulation, the \emph{Attack Orchestrator} allows the system to monitor real-time security risks and effectively validate mitigation strategies. Additionally, new attack types can be easily added, ensuring the system remains relevant as new threats emerge.

\myparagraph{E. Database Module} The \emph{Database Module} is responsible for storing all the logs and data generated during the simulation. The system uses MongoDB\footnote{https://www.mongodb.com/}, a NoSQL database, which provides efficient storage and retrieval of structured and unstructured data. Key functionalities of this layer include Storing EV charging logs and system performance metrics. The NoSQL-based structure enables high-speed queries and flexible data storage, making it ideal for handling large-scale simulations while maintaining data integrity.

\myparagraph{F. Simulator Suite} The base layer of the orchestrator is the \emph{Simulator Suite}, which integrates multiple EV charging simulators and emulators to provide a robust testing environment. The current version supports:
\emph{(I) EVerest}\cite{everest}: A powerful open-source software stack that supports EV charging infrastructure testing.
(\emph {II) ISO 15118 Simulator}\cite{iso}: A dedicated simulator to test charging communication using the ISO 15118 protocol, ensuring compatibility with real-world implementations.
\emph{ (III) Mock Simulator}: A lightweight, configurable simulator designed for testing scenarios that do not require full-scale simulation complexity.
\emph{(IV) ACNSim}\cite{lee2019acnsim}: A simulator for modelling the behaviour of EV and EVSE charging.
The \emph{Simulator Suite} is modular in design, allowing new simulators to be easily added. This adaptability ensures that the system remains future-proof and can accommodate evolving EV charging technologies.

\begin{figure*}
    \centering
    \includegraphics[width=0.9\linewidth]{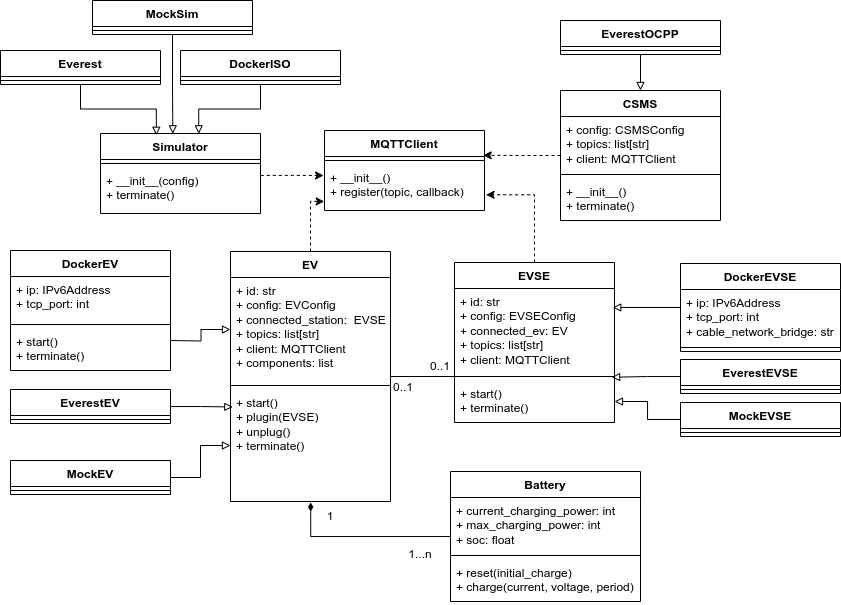}
    \caption{\projecttitle EV Orchestrator class diagram}
    \label{fig:class-diagram}
\end{figure*}

\subsection{\projecttitle Design and Implementation} \label{sec:design}
This section discusses the design and implementation of \projecttitle, including the class diagram of \emph{EV Orchestrator}, the charging sequence diagram of \emph{EV Orchestrator} and the integration of \emph{EV Orchestrator} with \emph{Attack Orchestrator} and \emph{Database}.

\subsubsection{Class Diagram of EV Orchestrator} \label{sec:class-diagram}
The class diagram shown in Figure \ref{fig:class-diagram} represents the core architecture of the \emph{EV Orchestrator}. The architecture is structured around three primary classes: EV, EVSE, and Simulator. These classes encapsulate the functionalities of an EV, an EVSE, and the simulation environment, respectively. At the centre of the figure, MQTT facilitates communication between these components, ensuring real-time data exchange.

The EV class models the behaviour of an EV within the simulation. It includes two primary functions that interact with the EVSE class: \emph{plug in} and \emph{plug out}. Additionally, each EV instance contains at least one instance of the Battery class, which simulates the behaviour of an EV battery during the charging process. The EVSE class is responsible for managing the charging session, handling power delivery, and modifying runtime variables such as current and voltage levels based on the simulation conditions.

The Simulator class serves as the orchestrator, responsible for configuring, deploying, and managing instances of different simulators. It abstracts the complexities of different simulation environments, providing a unified interface for handling various EV charging protocols and behaviours.

Each of these base classes, EV, EVSE, and Simulator, is further extended into subclasses to support different simulator implementations. In our current implementation, we have integrated three distinct simulators: \emph{MockSim}, \emph{Docker ISO}, and \emph{EVerest}. Correspondingly, each of the three main classes has dedicated subclasses for these simulators. EV class extends to MockEV, DockerEV, and EverestEV, each representing an EV model within its respective simulation environment. EVSE class extends to MockEVSE, DockerEVSE, and EverestEVSE, each defining the behaviour of an EVSE according to the underlying simulator. Finally, the Simulator class is extended to MockSim, DockerISO, and Everest, handling the deployment and execution of their respective simulation environments.

This modular and extensible design ensures that new simulators can be easily integrated into the framework by simply extending the base classes. The use of MQTT for inter-component communication enables seamless interaction between different simulator instances, making the system adaptable for future enhancements and real-world testing scenarios.

\subsubsection{Interaction of Attack Orchestrator and Database with EV Orchestrator}

    \begin{figure*}
        \centering
        \includegraphics[width=0.75\linewidth]{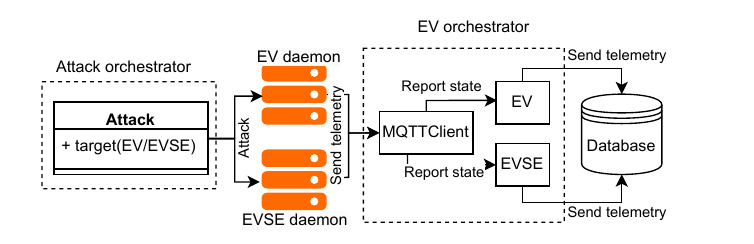}
        \caption{Interaction between \projecttitle components}
        \label{fig:orchestration-integration}
    \end{figure*}

 The integration between the EV orchestrator, Attack orchestrator, and Database involves multiple stages of communication as shown in Figure \ref{fig:orchestration-integration}. The EV daemon and the EVSE daemon represent the simulation suite, executing the charging process. The Attack orchestrator performs the attack directly on the EV and EVSE daemons. As the attack continues, the MQTT client inside the EV orchestrator collects the telemetry from the daemons and invokes the instances of EV and EVSE. These instances will update their internal state representation based on the telemetry of the daemons, as well as record the telemetry data to the database. The Database serves as historical storage for charging data metrics, allowing for post-simulation analysis. 

\section{E\lowercase{valuation}}
\label{sec:evaluation}

This section evaluates \projecttitle for two separate types of attacks (A) Broken wire attack and (B) Fuzzification attack. The detailed implementation and analysis are given below.

\subsection{Case 1: Broken wire attack}\label{sec:case1}
A Broken wire attack in an EV charging system occurs when an attacker deliberately manipulates the communication or power transfer wires between the EV and EVSE~\cite{kohler2022brokenwire}. This can disrupt charging sessions, cause failures in power delivery, or even trick the system into falsely detecting a disconnected vehicle. If the control signal is compromised, the EVSE may be unable to regulate power flow properly, leading to potential safety hazards or denial-of-service (DoS) attacks. 

Table~\ref{tab:bk-scope} categorises broken-wire attack scenarios on EV charging infrastructure based on their scope, impact, and attack methods. Individual-targeted attacks primarily exploit the physical communication layer (L1), specifically the ISO~15118 protocol, to disable a single charger.  
This prevents an EV from charging, causing inconvenience to the driver, but with limited overall impact. Group-level attacks escalate the disruption by targeting the network layer (L3), potentially affecting multiple chargers in a given area. 
By interrupting communication between the chargers and the energy management system, these attacks can make EVSEs in a region unavailable, forcing drivers to seek alternative locations and creating bottlenecks. Large-scale attacks pose the most severe risk, potentially disrupting all chargers across an entire city. By manipulating vulnerabilities at multiple layers, these attacks could disable widespread charging infrastructure, leading to significant traffic congestion and power distribution challenges. While the specific methods for large-scale attacks are not always detailed, the table highlights how different layers of the EV charging system can be exploited.

\begin{table*}
    \small
    \centering
    \caption{Disruptions caused by the broken wire attack}
    \resizebox{\textwidth}{!}{
    \begin{tabular}{|>{\centering\arraybackslash}p{0.15\linewidth}|>{\centering\arraybackslash}p{0.15\linewidth}|>{\centering\arraybackslash}p{0.15\linewidth}|>{\centering\arraybackslash}p{0.15\linewidth}|>{\raggedright\arraybackslash}p{0.05\linewidth}|>{\centering\arraybackslash}p{0.15\linewidth}|>{\raggedright\arraybackslash}p{0.15\linewidth}|} \hline 
         \textbf{Scope of the attack}&  \textbf{Chargers availability}&  \textbf{Outcome of the attack}&  \textbf{Attack mimicking technique} &\textbf{Layer}& \textbf{Result} &\textbf{Effect}\\ \hline 
         1 driver&  Limited number of chargers unavailable&  Frustration of the driver as it impacts daily schedule&  Attack on the physical network layer between EV and the EVSE &L1, L2& Disruption in the ISO15118 communication protocol; communication error raised &Charging process stopped; charger state switched to mated\\ \hline 
         Group of the drivers&  Chargers unavailable in a region&  Bottlenecks at the EVSEs as users try different EVSEs&  Error condition in the EVSEs &L3, L4&  Charging process stopped; reduction in the power consumed for charging the cars&Unavailability of EVSEs in the region\\ \hline 
         All drivers&  No chargers available&  Disruption in the city traffic&   -&-&  -&-\\ \hline
    \end{tabular}}
    \label{tab:bk-scope}
\end{table*}

The Attack Orchestrator generates the Broken Wire attack in two scenarios: \textit{(A) L1-L2 level} - The Pumba chaos engineering tool~\cite{pumba} is used to interrupt the EV-EVSE communication that disrupts the ISO 15118 protocol execution; \textit{(B) L3-L4 level} - The charging process is disrupted in the EV orchestrator to raise the error flag. 

\subsubsection{L1-L2 Level attack} 
\begin{figure}
    \centering
    \includegraphics[width=0.95\linewidth]{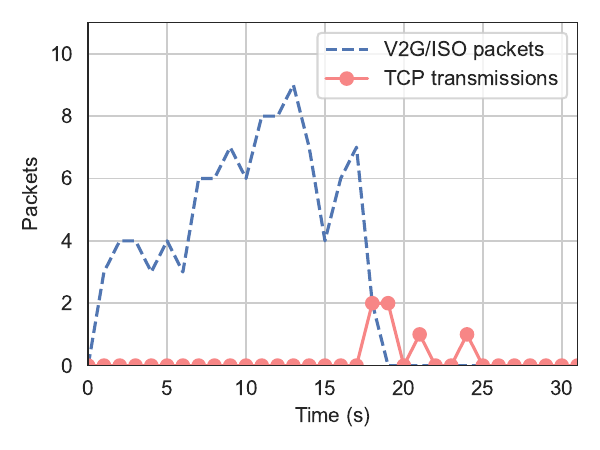}
    \caption{Packet capture during a Broken Wire attack}
    \label{fig:broken_wire}
\end{figure}

This scenario is modelled by containing a pair of EV and EVSE daemons connected to a separate network bridge that acts as the charging cable between the two. The EV daemon will establish a connection with the EVSE to initiate charging. Pumba internally uses the Linux tool \emph{tc} for network traffic control to manipulate any network interfaces of a machine. For the L1-L2 attack, Pumba emulates packet loss on the connection bridge that the EV and EVSE daemon are using, which effectively simulates the act of severing the physical communication wire. After the attack is initiated, all communications between the two actors are lost, and the charging session should abort as the ISO 15118 protocol requires heartbeat messages to be exchanged during the charging loop. Figure \ref{fig:broken_wire} shows the packet count during the evaluation. During the attack, starting from the $18$th second, TCP error packets have been observed (\textit{TCP transmissions}),
which is indicated by the red lines.
The amount of time to terminate a charging session depends on which state the charging session is currently in. In this case, the EVSE was in Charge state and needs to respond with a \texttt{ChargingStatusRes} on ISO 15118 message to the EV during the charging loop. The timeout window for that state is $2$ seconds, after which the session terminates without any message exchanges. 


\subsubsection{L3-L4 Level attack}
\begin{figure}
    \centering
    \includegraphics[width=0.95\linewidth]{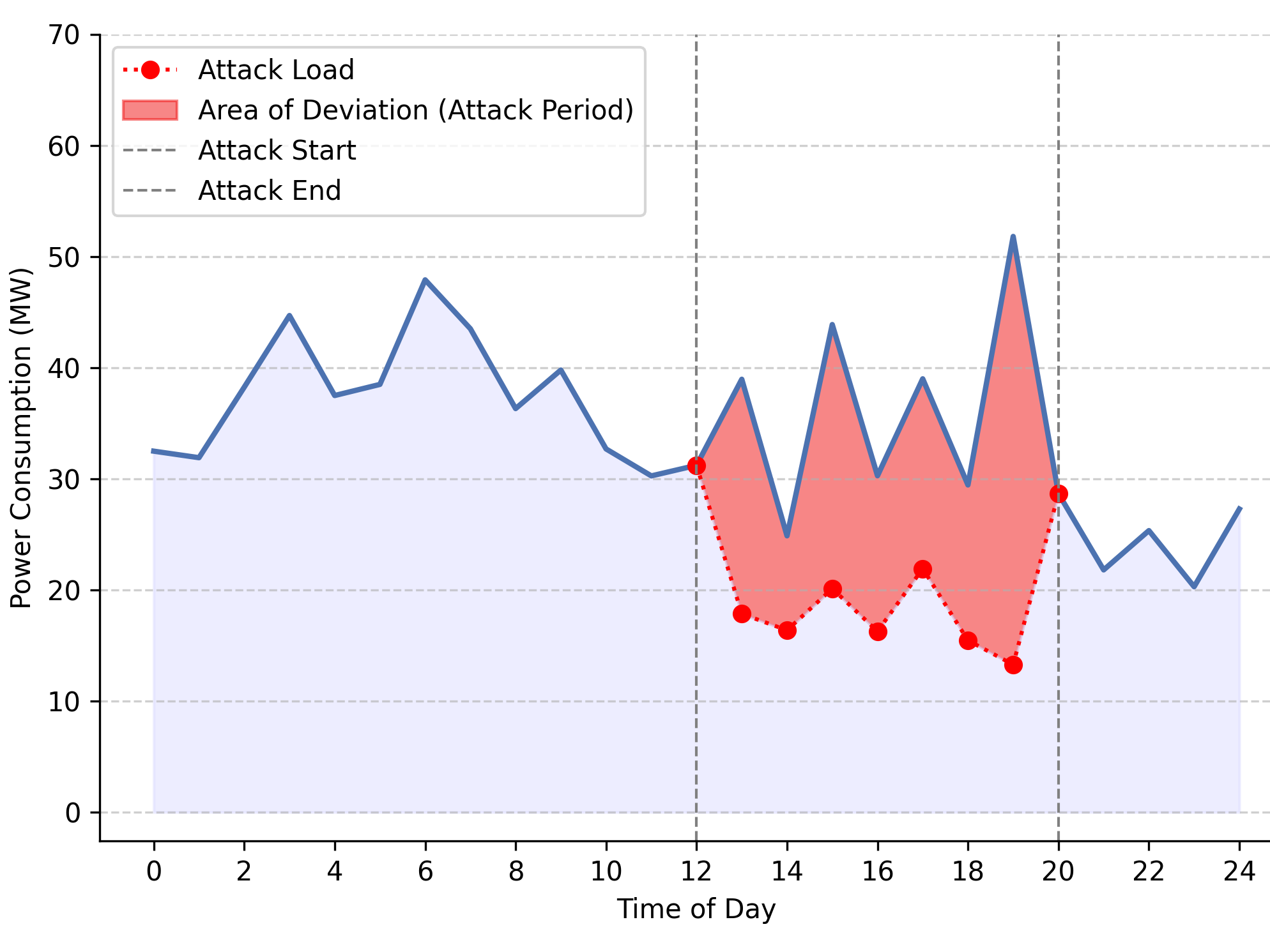}
    \caption{Power Consumption Pattern at EVSE: Impact of Broken Wire Attack Showing Deviation from Normal Load (MW) over 24-Hour Period}
    \label{fig:broken_wire_power}
\end{figure}

The effects of a broken wire attack on an EVSE’s power consumption over a 24-hour period are illustrated in Figure \ref{fig:broken_wire_power}. Here, the blue line represents the expected (normal) power consumption pattern, while the red dotted line indicates the compromised load factor during the attack period, which occurs between the two vertical dashed lines (approximately from hour 12 to hour 20). During this attack window, a significant deviation in power consumption is observed, as highlighted by the red shaded area. The actual power delivered drops well below the expected levels, showing a fluctuation in the range of 15-20 MW, whereas under normal conditions, the power demand in this timeframe was expected to be around 30-40 MW. This results in an estimated 40-50\% reduction in power delivery, significantly affecting EVSE performance.

Such a pattern is characteristic of a broken wire attack, where a deliberate disruption in the charging infrastructure leads to a sharp decline in power utilisation. This reduction in available power can cause severe charging delays, bottlenecks at stations, and user inconvenience, particularly in high-demand scenarios. In a larger context, such attacks could also strain the energy grid by causing unexpected demand shifts, leading to load imbalances and operational inefficiencies.

\subsection{Fuzzification attacks}\label{sec:case2}
This section implements and evaluates a fuzzification attack targeting the OCPP protocol implementation within the EV charging ecosystem. Given the increasing reliance on OCPP for secure and reliable communication between CS and Central Management Systems (CSMS), it is crucial to assess its resilience against protocol fuzzing. Our approach systematically evaluates a generic OCPP $2.0.1$ implementation using the \projecttitle, with a focus on identifying security weaknesses through protocol-based fuzzing at the L2 level. While our work builds on OCPPStorm~\cite{coppoletta2024ocppstorm}, the key distinction lies in leveraging our \emph{Attack orchestrator} of \projecttitle to conduct a fuzzification attack and evaluate the implementation.

The \emph{Attack orchestrator} initiates the fuzzing process by systematically sending OCPP requests (with message code $2$) while recording the backend responses. The key components of the attack are \emph{A) Fuzzer}, which operates independently of the orchestrator and injects malformed or unexpected OCPP messages. \emph{B) Orchestrator Attack Handler}, which manages attack execution and interacts with the Language-Based Fuzzer. \emph{C) Message Validator} to ensure the correctness of messages before they are processed by the backend. For this attack, we assume that the fuzzer operates as a malicious EVSE, either bypassing authentication layers or functioning in a controlled test environment where security constraints are deliberately removed. 
For the implementation, we utilize the ISLa fuzzer~\cite{10.1145/3540250.3549139}, which applies grammar-based fuzzing techniques to explore vulnerabilities in OCPP message exchanges. ISLa's capabilities in generating context-sensitive inputs align with the complexities of OCPP messages. 
Our attack strategy specifically targets the robustness of the OCPP backend (Citrine-OS) within the EVerest simulator stack. To ensure the validity of the generated messages, we integrate the typed-ocpp library\footnote{https://github.com/jacoscaz/typed-ocpp}, which verifies compliance with the OCPP standard.

\subsubsection{Fuzzification Attack Results and Discussions}
\begin{figure*}[t]
    \centering
    \includegraphics[width=0.8\textwidth]{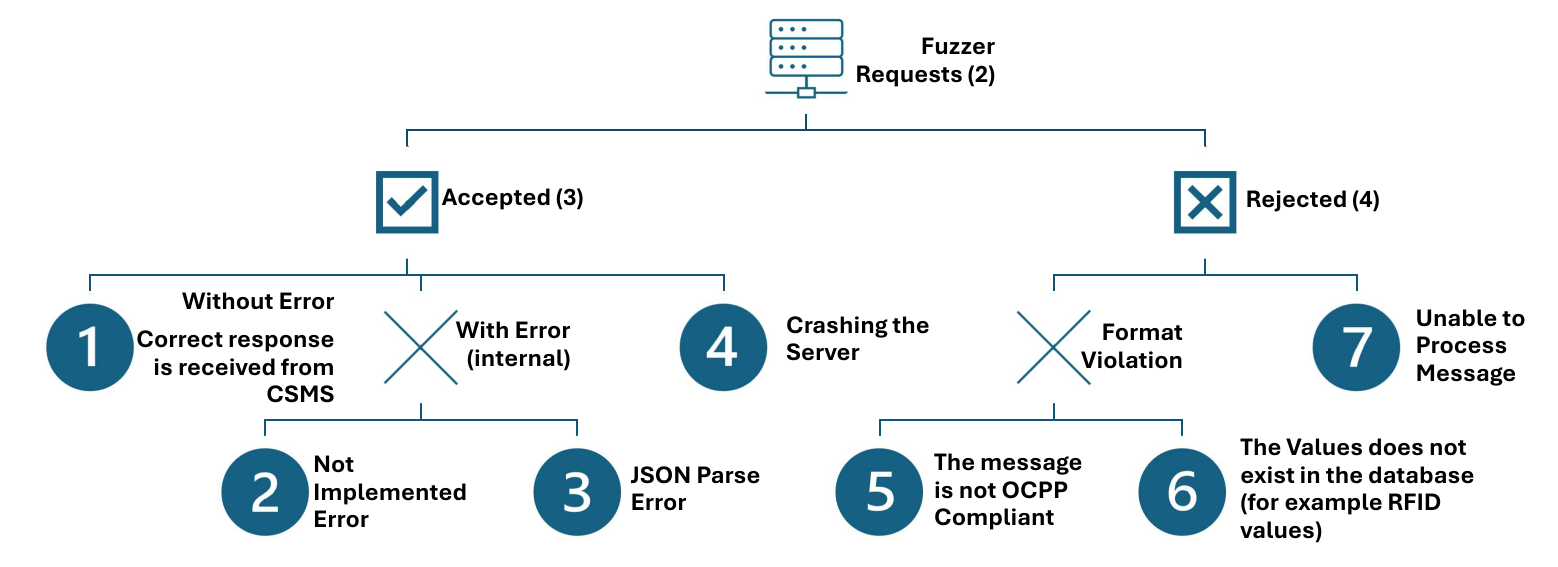}
    \caption{Possible Outcome Scenarios for Fuzzification}
    \label{fig:fuzz-possible-outcome}
\end{figure*}

\subsubsection{Fuzzification Strategy}
The attack consists of two primary fuzzing strategies, as discussed below.

\myparagraph{Randomised Message Fuzzing (Language-based)} This strategy involves sending a randomised sequence of OCPP messages without following a structured transaction flow. Each of the $10$ most commonly used OCPP messages is sent $100$ times in an unordered manner. The objective is to test how the backend handles unexpected message sequences and excessive load against the normal protocol behaviour.

This approach is particularly useful for a) stress-testing the backend by sending repeated requests within a short timeframe, b) identifying inconsistencies in response handling, where the backend might incorrectly process messages outside the expected workflow, and c) evaluating resilience against high-frequency malformed requests and understanding the system’s rate-limiting mechanisms. For instance, an \emph{Authorise request} message may fail if the backend does not recognise the RFID key due to the absence of prior \emph{BootNotification} or \emph{Registration} steps. Although somewhat forced, this method allows us to analyse the system’s resilience to unexpected message patterns and its ability to handle stress under randomised loads.  

\myparagraph{State-Based Language Fuzzification} This approach follows structured communication flows to assess how the backend processes logically ordered transactions. Messages are categorised into four functional groups: \emph{A) Start-up Phase Messages}, e.g., \texttt{BootNotification}, \texttt{FirmwareStatusNoti-} \texttt{fication}, and \texttt{ClearCacheRequest}, \emph{B) Operational Phase Messages}, e.g., \texttt{Heartbeat}, \texttt{StatusNotificationRequest}, \emph{C) User Interaction Messages}, e.g., \texttt{AuthorizeRequest}, \texttt{Get15118} \texttt{EV-CertificateRequest}, \texttt{NotifyCustomerInformation}, and \emph{D) Firmware \& Custom Data Handling}, e.g., \texttt{PublishFirmwareSt-} \texttt{atusNotificationRequest}, \texttt{DataTransferRequest}. A real behaviour of EV charging is represented by executing these messages in the correct sequence. Each sequence of messages is executed $100$ times to assess system consistency. The fuzzer begins with a valid initial sequence but deliberately injects errors or malformed messages at specific transition points. The aim is to analyse how the backend processes unexpected sequences and whether it can recover gracefully from erroneous
transactions. This method helps uncover protocol weaknesses related to state transitions and error recovery mechanisms, data validation failures, and potential bypass scenarios.

Based on our observations, the messages can either be \emph{Accepted} or \emph{Rejected}, as shown in Figure \ref{fig:fuzz-possible-outcome}. If CSMS accepts the message without any error, the return code is \texttt{(3,1)} {\Large\textcircled{\small 1}}. The message can also be accepted with an erro,r generating the return code as \texttt{(3,2)} and \texttt{(3,3)} for Not Implemented Error {\Large\textcircled{\small 2}} and JSON Parse Error {\Large\textcircled{\small 3}}, respectively. The request may get accepted, crashing the server with a return code \texttt{(3,4)} {\Large\textcircled{\small 4}}. Moreover, some messages were also rejected. The message can be rejected due to non-compliant format violation {\Large\textcircled{\small 5}} or non-existent value {\Large\textcircled{\small 6}}. Finally, if the CSMS is not able to process some messages, they are rejected with an error code of \texttt{(4,7)} {\Large\textcircled{\small 7}}.


The results of the fuzzification attacks are summarised in Table \ref{tab:fuzzification}. This table compares randomised message fuzzification with state-based fuzzification. The observed responses provide insights into how the backend processes messages under various fuzzing conditions.

In random fuzzification, where each message was sent independently $100$ times, certain messages, such as \textit{Heartbeat} and \textit{FirmwareStatusNotification}, were consistently accepted. This suggests the independence of such messages on prior states or primarily function as boot messages. In contrast, \textit{AuthorizeReq} had a $78$\% acceptance rate, with $19$\% intermediate responses and $3$\% failures. This variability is likely due to the absence of a valid RFID token in some instances. Since \textit{BootNotification}, such as registering an EVSE, must precede other messages in a typical OCPP transaction, it consistently caused the server to stop, likely for security reasons. Similarly, \textit{ClearCacheReq} was mostly rejected, reinforcing that the backend enforces security mechanisms against unauthorised cache manipulations. \textit{DataTransferReq} was fully accepted in all cases, indicating potentially weaker validation in the backend. Meanwhile, \textit{Get15118EV-CertificateReq} exhibited high latency ($573$ ms), likely due to the cryptographic overhead of certificate verification or the search for certificates in the backend or databases.

\begin{table*}[htbp]
  \micro
   \centering
    \caption{Fuzzification attack analysis in \projecttitle}
    \resizebox{1.85\columnwidth}{!}{  
    \begin{tabular}{|p{2cm}|p{1.5cm}|p{1.5cm}|p{1.5cm}|p{1.5cm}|p{1.05cm}|p{1.5cm}|p{1.5cm}|p{1.05cm}|p{1.05cm}|}
    \hline
    \textbf{Message} & \textbf{Accepted (code 3, 1)}   & \textbf{Accepted (code 3, 2)}  & \textbf{Accepted (code 3, 3)}  & \textbf{Stopping the server (code 3, 4)} & \textbf{Rejected (code 4, 5)}  & \textbf{Rejected (code 4, 6)} & \textbf{Rejected (code 4, 7)}  & \textbf{Average Latency} \\ \hline
    Heartbeat   & 100\%   & 0   & 0   & 0   & 0   & 0   & 0  & 10ms  \\ \hline
    AuthorizeReq  & 78\%  & 19\%  & 3\%  & 0  & 0  & 0  & 0  & 11ms   \\ \hline
    BootNotification  & 0  & 0  & 0  & 100\%  & 0  & 0  & 0  & 48ms    \\ \hline
     ClearCacheReq  & 0  & 0  & 0  & 0  & 92\%  & 0  & 8\%  & 3ms    \\ \hline
       FirmwareStatus-
       Notification  & 100\%  & 0  & 0  & 0  & 0  & 0  & 0 & 26ms    \\ \hline
        DataTransfer-Req  & 0  & 100\%  & 0  & 0  & 0  & 0  & 0 & 8ms    \\ \hline
        Get15118EV-CertificateReq  & 0  & 0  & 100\%  & 0  & 0  & 0  & 0 & 573ms    \\ \hline
        NotifyCustomer-Information  & 100\%  & 0  & 100\%  & 0  & 0  & 0  & 0 & 8ms    \\ \hline
         StatusNoti-ficationReq  & 100\%  & 0  & 100\%  & 0  & 0  & 0  & 0 & 16ms    \\ \hline
         PublishFirm-wareStatusNot-ificationReq
 & 0  & 34\%  & 0  & 0  & 66\%  & 0  & 0 & 10ms    \\ \hline
    \end{tabular}
    }
    \label{tab:fuzzification}
\end{table*}

In state-based fuzzification, messages were sent in structured sequences, leading to improved system response accuracy. \textit{AuthorizeReq} exhibited a higher success rate, confirming that adherence to the correct transaction flow enhances acceptance. \textit{NotifyCustomerInformation} and \textit{StatusNotificationReq}, which had fluctuating responses in randomised tests, demonstrated greater consistency when executed in proper sequences, suggesting their reliance on prior information. Similarly, \textit{PublishFirmwareStatusNotificationReq}, which produced mixed results in random fuzzing, followed a more predictable pattern when sent in structured sequences.

These findings suggest that \projecttitle can generate various fuzzification attack-based tests to reveal how the backend processes unstructured traffic. 
\section{D\lowercase{iscussion}}
\label{sec:discussion}

While \projecttitle has demonstrated its ability to orchestrate and analyse cyber threats in the EV charging ecosystem, several enhancements can further strengthen its capabilities.

\myparagraph{Integration of Additional EV Simulators} Expanding \projecttitle to support a wider range of EV simulators, including proprietary and hardware-in-the-loop simulators, will improve its applicability across different charging environments. This will enable more realistic testing scenarios and broader compatibility with real-world EV infrastructure.

\myparagraph{Incorporation of Additional Attack Types} While the current implementation of \emph{Attack Orchestrator} focuses on Broken Wire and Fuzzification attacks, future iterations will incorporate a diverse range of cyber threats, such as relay attacks, Man-in-the-Middle attacks, and side-channel attacks on charging communication protocols. This will provide a comprehensive security assessment of vulnerabilities across multiple layers of the charging ecosystem.

\myparagraph{Development of a User Interface for \projecttitle}
We are working on developing a graphical user interface (GUI) for \projecttitle to enhance usability, allowing security researchers and infrastructure operators to configure attack scenarios, monitor real-time simulation data, and visualise results more intuitively. This will improve accessibility and ease of deployment for broader adoption in industry and academia.

\myparagraph{Predictive Model for Future Attack Prediction} 
Leveraging the collected data from the attack simulations, machine learning models will be developed to predict potential (future) cyberattacks on EV charging infrastructure. By analysing historical attack patterns and system behavior, \projecttitle could provide early warnings and proactive defense mechanisms, reducing the risk of large-scale disruptions.
\section{Conclusion}
\label{sec:conclusion}

A multi-layered orchestrator, \projecttitle, for the EV charging ecosystem, which integrates various existing EV simulators is described. This orchestrator also supports the execution of cyberattacks on the EV ecosystem, and a database to systematically evaluate threats in EV charging systems.  
We evaluated \projecttitle using two different attack scenarios: Broken Wire attack (a cyber-physical attack) and a frame fuzzification attack (a cyberattack), generated using the Attack Orchestrator. The results demonstrate that \projecttitle can seamlessly orchestrate and execute these attacks while accurately monitoring their effects on the EV ecosystem, offering valuable insights into potential vulnerabilities and provide the basis to improve system resilience. Moreover, the framework's modularity allows for the integration of new attack models, security policies, and countermeasures, making it a powerful tool for ongoing research in EV security and threat mitigation.

\section*{Acknowledgment}
This study was partly supported by the National Edge AI Hub for Real Data: Edge Intelligence for Cyberdisturbances and Data Quality (UKRI EPSRC EP/Y028813/1) and EPSRC UK-Australia Centre in a Secure Internet of Energy: Supporting Electric Vehicle Infrastructure at the ``Edge'' of the Grid (EPSRC, EP/W003325/1).

{
\bibliographystyle{IEEEtran}
\bibliography{TC/main_ref}   
\section*{Appendix I: EV Orchestrator State Transition in \projecttitle }
\begin{figure*}[t]
    \centering
    \includegraphics[width=0.7\linewidth]{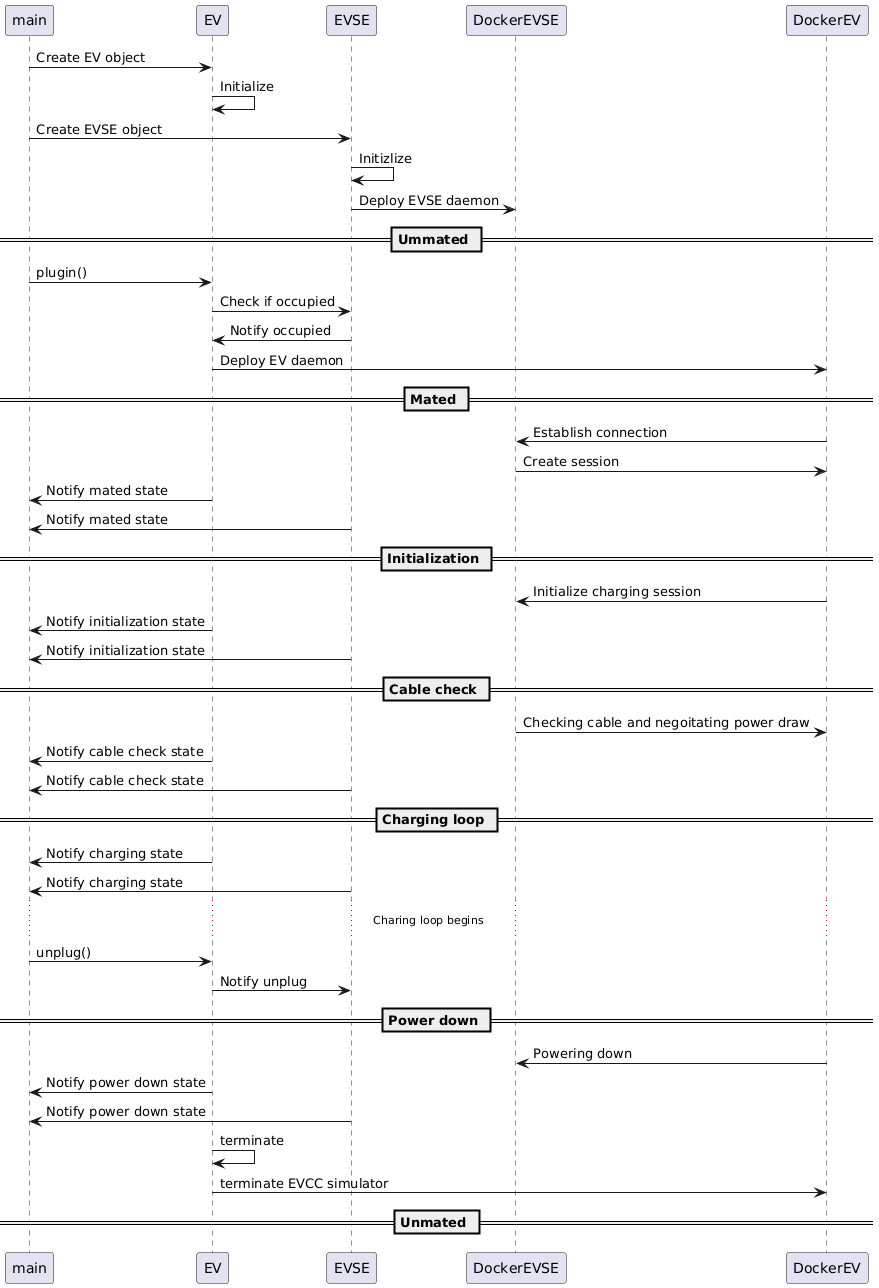}
    \caption{Sequence diagram showing normal charging process in \projecttitle}
    \label{fig:seq-diagram}
\end{figure*}

The diagram in Figure \ref{fig:seq-diagram} shows the series of interactions involved in the EV charging process in the EV Orchestrator of \projecttitle. The simulation begins with the user specifying a predefined charging scenario, defining the configuration of various components, including EV type, EVSE type, charge required and EVSE charge rate. It also allows a user to specify any conditional triggers that affect the charging system's behaviour.

Given the configuration, the Simulator first creates and initialises EV and EVSE objects. Depending on the Simulator specified in the configuration, respective instances of EV and EVSE daemons will be created. Figure \ref{fig:seq-diagram} shows the interaction for the Docker ISO simulator. Initially, the EVSE is Unmated. The EV checks the state of EVSE, and if it is unoccupied, an EV daemon is deployed. Next, communication between the EV and EVSE daemons will be established, and the charging session will be initialised. Following that, a compulsory cable check operation will be performed, and if that is successful, charging will start. The charging will finish when either the EV battery is charged to 100 \% or a user initiates an interrupt. This leads to an unplug event, and the EVSE daemon will start powering down. The simulator is notified of that operation, which terminates the EV daemon. After the process finishes, all the system logs are collected and stored in the database for further processing.

By leveraging the modular architecture described in the class diagram, the system ensures flexibility in testing various charging scenarios and protocols, supporting both software-based simulations and real-world EVSE interactions.
}
 
\end{document}